6
13
14
14
1

# A multicolor survey of absolute proper motions : Galactic structure and kinematics in the direction of galactic center at intermediate latitude*

D.K.Ojha[1,2], O.Bienaymé[1], A.C.Robin[1,2], and V.Mohan[3]

[1] Observatoire de Besançon, 41 bis, Av de l'Observatoire, BP1615, F-25010 Besançon Cedex, France
[2] Observatoire de Strasbourg, CNRS URA 1280, 11 rue de l'Université, F-67000 Strasbourg, France
[3] Uttar Pradesh State Observatory, Manora Peak, Nainital, 263129, India

**Abstract.** Bienaymé et al. (1992) have published a *magnitude-limited* sample of proper motions for stars at the intermediate galactic latitude ($l = 3°$, $b = 47°$; $\alpha_{1950} = 15^h 18^m$, $\delta_{1950} = +02°16'$) for a 2 square degree field. In their study there was a lack of an accurate absolute astrometric reference due to very small number of galaxies or extragalactic objects in the field. However, since Bienaymé et al. paper, more deep photographic plates and CCD standards have been obtained. Here we have derived a new photographic photometry and proper motions for ∼ 20000 stars with completeness to V ∼ 18 in the same direction for a 15.5 square degree field. The combination of four glass copies of the Palomar Observatory Sky Survey (i.e. POSS 1402 E & O and POSS 1429 E & O) has been used as a first epoch for proper motion determination. The random error of the proper motions is approximately 0."3 century$^{-1}$ to V∼17. The photometric accuracy ranges between 0.07 to 0.10 in the V, B and U bands. We stress the importance of the magnitude and color effects in astrometric surveys of field stars. Using color-magnitude diagrams of a few cluster member stars, a new distance of 6.9±0.5 kpc is derived for M5 and 20.3±0.8 kpc for Pal 5 globular clusters. This is in good agreement with other determinations.

We have analyzed the components of U+W and V galactic space motions resulting from the accurate proper motions survey. No dependence with z distance is found in the asymmetric drift of the thick disk population. New estimates of the parameters of the velocity ellipsoid have been derived for the thin disk, thick disk and halo populations of the Galaxy.

## 1. Introduction

With the advent of fast automatic measuring machines e.g. MAMA, COSMOS etc., it is now practicable to measure large number of stars recorded on the plates taken with powerful Schmidt telescopes. Bienaymé et al. (1992), Soubiran (1992a,b) and Ojha et al. (1994) have presented the combination of the OCA (Observatoire de la Côte d'Azur), ESO , Tautenburg, Palomar Schmidt plates and the MAMA automated-computerized plate scanner to produce a large survey for proper motions down to the limiting magnitude of the Schmidt plates in 3 different directions (galactic center, north pole and galactic anticenter) of the meridian plane. The accuracy in all the above surveys ranges between 0."2 to 0."4 century$^{-1}$.

One of the topics rediscussed in recent years is the question whether the observed data require an intermediate field star component (called it a thick disk) in addition to an exponential disk, with scale height of ∼ 300 pc and a spheroid. Gilmore and Reid (1983) have suggested that the Galaxy possesses an intermediate population component in addition to its well-established (thin) disk and halo populations. They argued that deep starcounts could not be modeled by any plausible two-component system. Recently much data have been presented advocating for such a component. One of the remarkable points is that the existence of such a component (thick disk) follows not only from photometric data (see Fenkart 1988) but also from kinematic data (Sandage 1987; Norris 1986; Wyse & Gilmore 1986; Freeman 1987; Bienaymé et al. 1990; Soubiran 1993; Ojha et al. 1994).

Here we describe a new multicolor survey of absolute proper motions at the intermediate galactic latitude field ($l = 3°$, $b = 47°$; $\alpha_{1950} = 15^h 18^m$, $\delta_{1950} = +02°16'$). In this study new proper motion data in combination with photometric data are used to derive some properties of the vertical structure of the Galaxy, focussing on the kinematic properties of the intermediate component.

The outline of the paper is as follows: section 2 summarises the photographic plates used in the present survey. In sections 3 and 4, we present the technical aspects of the photometric and astrometric measurements to produce accurate data sets. We also discuss the various problems in calibrating Schmidt plates and how to correct the major effects. Section 5 reviews the comparison of our photometric and proper motion data with the other data sets. In section 6, we discuss the implications of these data to investigate the structure and kinematics of the Galaxy.

## 2. Plate material

Details of the plates used in the present discussion are given in Table 1. The plates were measured with the MAMA measuring machine in Paris. The combination of Palomar and ESO Schmidt plates (1955-1989) has been used for proper motions determination giving a time base of $\sim$ 34 years. OCA and ESO plates were used for photometric purposes.

## 3. Photographic photometry

Photoelectric and CCD standard stars from the following sources in U, B and V bands were used to calibrate the Schmidt plates: Arp (1962), Sandage & Hartwick (1977), Smith et al. (1986), Richer & Fahlman (1987), Lasker et al. (1988), Stetson & Harris (1988), Cayrel (1993), Rees (1993), and Mohan (1987b,1991,1994). Altogether, 200 standards are available in B and V bands, however, they are poorly distributed over the field in case of the U band. Result of the calibration using a fourth-order polynomial for the ESO Schmidt plate 7958 is shown in Fig. 1. The following corrections depending on color were applied on the V & B ESO and U OCA Schmidt plates:

$v_{inst} = V_{Johnson} - 0.118(B - V)$

$b_{inst} = B_{Johnson} - 0.113(B - V)$

$u_{inst} = U_{Johnson} - 0.155(B - V) + 0.096(U - B)$

The color coefficients for the V and B ESO Schmidt plates were derived by using standards well distributed over the entire field and whose positions were well determined (for general treatment see Mohan & Crézé 1987a). An iterative process was used to improve upon the coefficients. The color equation for the OCA plate in U band is taken from Mohan & Crézé (1987a).

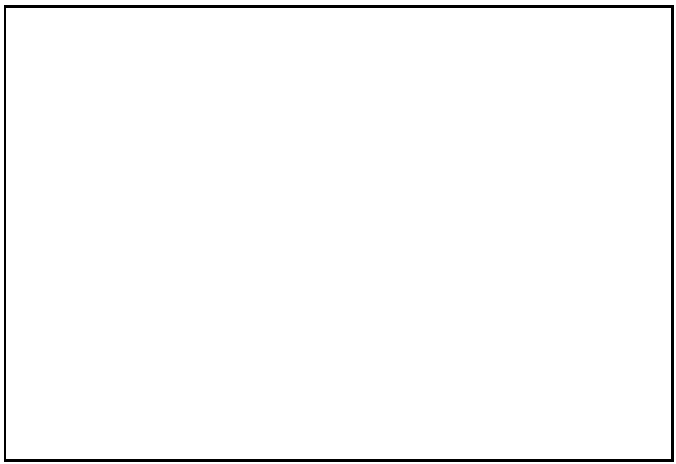

Fig. 1. Calibration curve from $\sim$ 200 standards for ESO 7958 Schmidt plate. The dispersion is 0.07 mag

In the photometric reduction of the whole Schmidt plate, the main problem arises from the geometrical effect i.e the variation of the instrumental response as a function of position in the field. We have corrected this effect in the following way :

The bright standards well distributed over the entire field have been used for this purpose. Using a linear fit in $x$ and $y$ coordinates, the geometrical coefficients have been determined for each filter/emulsion combination. These corrections have been applied in V, B and U magnitudes for all the stars on the plates. This correction is sufficient to minimize the geometrical variation between two plates in same filter/emulsion combination ( the difference of magnitudes between two plates does not depend on position). However, the plate to plate magnitude comparison between two plates in different filter/emulsion combination still shows the geometrical variations or systematic field effects due to sensitivity variations according to position. This effect can be seen in B-V and U-B color indices (the maximum variations of $\pm 0.12$ and $\pm 0.05$ on the mean U-B and B-V color indices between subfields respectively).

To correct this effect, we have applied another correction on B-V and U-B color indices called "flat fielding" described by Bienaymé et al. (1992). This correction can be estimated if the average color of stars in any subfield is position independent. This is true since no differential absorption effects are expected in this intermediate latitude field. Using a polynomial fit in $x$ and $y$ coordinates, we have corrected the B-V and U-B indices so that their mean values for each subfield remain constant over the whole field. The correction shifts indices by an unknown constant which is determined using the bright standards.

A relation between logarithm of integrated flux measured by MAMA and object area (Ojha et al. 1994) has been used to separate the resolved objects (e.g. galaxies, blends etc.) from the unresolved ones (e.g. stars). A deep ESO 7992 plate in V filter was used for this purpose. After stars/galaxies separation, our final proper motion catalogues having U, B & V magnitudes contain $\sim$ 12000 stars in a 15.17 square degree area, while $\sim$ 20000 stars have only B & V magnitudes in a 15.5 square degree area. The catalogues are complete down to V=18, B=19 and U=18.

The plate to plate dispersion in all magnitude ranges is shown in Table 2.

## 4. Astrometry

For the astrometric reduction, we have excluded the zone of the globular cluster M5. The zone (0.28 square degrees) is defined by cross-matching the cluster member stars measured precisely by Rees (1993) and our catalogue. The absolute proper motions of the field stars have been determined with respect to background galaxies. In matching the data sets from two plates, we have searched for all possible pairs within a radius of 40 $\mu m$ – corresponding to a proper-motion limit of $\sim$ 2.7 arcseconds for the ESO plate. Orthogonal functions were used to model the transform between the plate coordinates. The method used for the astrometric reduction is described in detail by Bienaymé et al. (1992), Bienaymé (1993) and Ojha et al. (1994).

We must point out here that the coordinates of the center of ESO and Palomar Schmidt plates are not the same. To cover the whole area of the ESO plate in astrometric reduction, we have used the combination of four POSS glass copies (i.e POSS 1402 O & E and POSS 1429 O & E) as a first epoch. The ESO 7992 plate was used as a reference plate for the astrometric reduction.

**Table 1.** Plate material

| Plate number | Emulsion+filter | Color | Exposure time(min) | Epoch | Scale ("/mm) |
|---|---|---|---|---|---|
| *OCA Schmidt plates* | | | | | |
| 500 | IIaO+UG1 | U | 120 | 06/05/1981 | 65.25 |
| 2446 | IIaO+UG1 | U | 120 | 04/03/1990 | 65.25 |
| *ESO Schmidt plates* | | | | | |
| 7958 | IIaO+GG385 | B | 60 | 04/04/1989 | 67.13 |
| 7970 | 103aD+GG495 | V | 60 | 06/04/1989 | 67.13 |
| 7992 | 103aD+GG495 | V | 60 | 10/04/1989 | 67.13 |
| 8003 | IIaO+GG385 | B | 60 | 12/04/1989 | 67.13 |
| *Palomar Schmidt plates* | | | | | |
| POSS 1402 | 103a-E+red plexiglass | red | 50 | 19/04/1955 | 67.13 |
| POSS 1429 | 103a-E+red plexiglass | red | 50 | 18/05/1955 | 67.13 |
| POSS 1402 | 103a-O+blue plexiglass | blue | 50 | 19/04/1955 | 67.13 |
| POSS 1429 | 103a-O+blue plexiglass | blue | 50 | 18/05/1955 | 67.13 |

**Table 2.** Dispersion of magnitudes from plate to plate comparison

| V | $\sigma_V$ | B | $\sigma_B$ | U | $\sigma_U$ |
|---|---|---|---|---|---|
| 10.25 | 0.04 | 10.25 | 0.03 | 10.25 | 0.04 |
| 10.75 | 0.04 | 10.75 | 0.04 | 10.75 | 0.06 |
| 11.25 | 0.05 | 11.25 | 0.05 | 11.25 | 0.10 |
| 11.75 | 0.05 | 11.75 | 0.04 | 11.75 | 0.08 |
| 12.25 | 0.06 | 12.25 | 0.05 | 12.25 | 0.10 |
| 12.75 | 0.05 | 12.75 | 0.05 | 12.75 | 0.10 |
| 13.25 | 0.05 | 13.25 | 0.06 | 13.25 | 0.10 |
| 13.75 | 0.05 | 13.75 | 0.05 | 13.75 | 0.10 |
| 14.25 | 0.06 | 14.25 | 0.05 | 14.25 | 0.10 |
| 14.75 | 0.06 | 14.75 | 0.06 | 14.75 | 0.10 |
| 15.25 | 0.06 | 15.25 | 0.06 | 15.25 | 0.09 |
| 15.75 | 0.07 | 15.75 | 0.06 | 15.75 | 0.08 |
| 16.25 | 0.07 | 16.25 | 0.06 | 16.25 | 0.08 |
| 16.75 | 0.08 | 16.75 | 0.06 | 16.75 | 0.08 |
| 17.25 | 0.08 | 17.25 | 0.07 | 17.25 | 0.08 |
| 17.75 | 0.09 | 17.75 | 0.07 | 17.75 | 0.09 |
| 18.25 | 0.11 | 18.25 | 0.07 | 18.25 | 0.08 |
| | | 18.75 | 0.08 | | |
| | | 19.25 | 0.08 | | |

*4.1. Plate to plate transform error*

The plate to plate transform error for 6th to 15th order $X$ and $Y$ transforms is about 0.4 $\mu m$ over most of the field between two ESO plates (ESO 8003 & 7992). Because of the better centering of the stellar images (see also Bienaymé 1993), the value of the transform error is very small. Figs 2 a & b show the difference between 15th and 6th order $X$ and $Y$ transforms in the two ESO plates. The amplitude of difference is about 1.2 $\mu m$, which is a small scale distortion in this particular type of reduction. However, in case of the Palomar and ESO plates (Figs 3ab), we can see large scale distortion of amplitude about 2.2 $\mu m$. The accuracy of the 6th to 15th order transform ranges from 1 and 2 $\mu m$ over most of the field and is not as good as in the previous example. This is because of the relative motions of the stars between two plates (mean motion of $\sim$ 0.5 $\mu m$) during 34 years. However, comparison between the 15th and 6th order $X$ and $Y$ transforms do not show any significant distortions in both examples. As a consequence, we have chosen the 6th order transform for our final astrometric reduction.

*4.2. Color effect on proper motions*

When comparing plates taken for the same field at different places, the main problem arises from the effect of differential atmospheric refraction, which causes relative displacement of images of stars of different colors. This systematic error induces a color term in the proper motions. This effect has been estimated using the equations given by Murray & Corben (1979) and Kovalevsky (1990). It can be seen in the Fig. 4 by Murray & Corben (1979) that for the V filter (GG495), the value of refraction constant (A) remains almost constant for different spectral types or wavelengths. However, for the B filter (GG385) the variation of A is important. Bienaymé et al. (1992) have calculated the expected color effect on the 2 plates (V and R filter) used for the astrometric reduction in 2 square degree field using the method given by Murray & Corben (1979) and Kovalvesky (1990) and found it small.

It is possible to minimize the color effect by using two plates in different filters centered on different wavelengths. As a consequence, for the ESO plates, we have applied the color correction on B filter with respect to V filter, assuming that the color term in V filter is almost zero in the following way :

As a first step, we have determined the proper motions ($\mu_x$ and $\mu_y$) of each star without color correction. In a second step, using a linear fit in the residuals of $x$ coordinate (parallel to $\delta$- coordinate) on the two ESO plates (V and B filters) *versus* B-V color, we have obtained the constants a1 & a2 for V filter and b1 & b2 for B filter as follows :

$$x_V = a1 + a2(B - V)$$

$$x_B = b1 + b2(B - V)$$

The bright stars in the magnitude range 12$\leq$V$\leq$16 were used for this purpose. The total color term in $x$ position is found to be 0.20 $\mu m$ (0.01 arcsecs) per unit in B-V (slope) for the B ESO plate. The following correction is applied in the $x$ coordinate of each star on the ESO plates :

$$x_{V\,cor.} = x_V \quad \text{(V filter)} \tag{1}$$

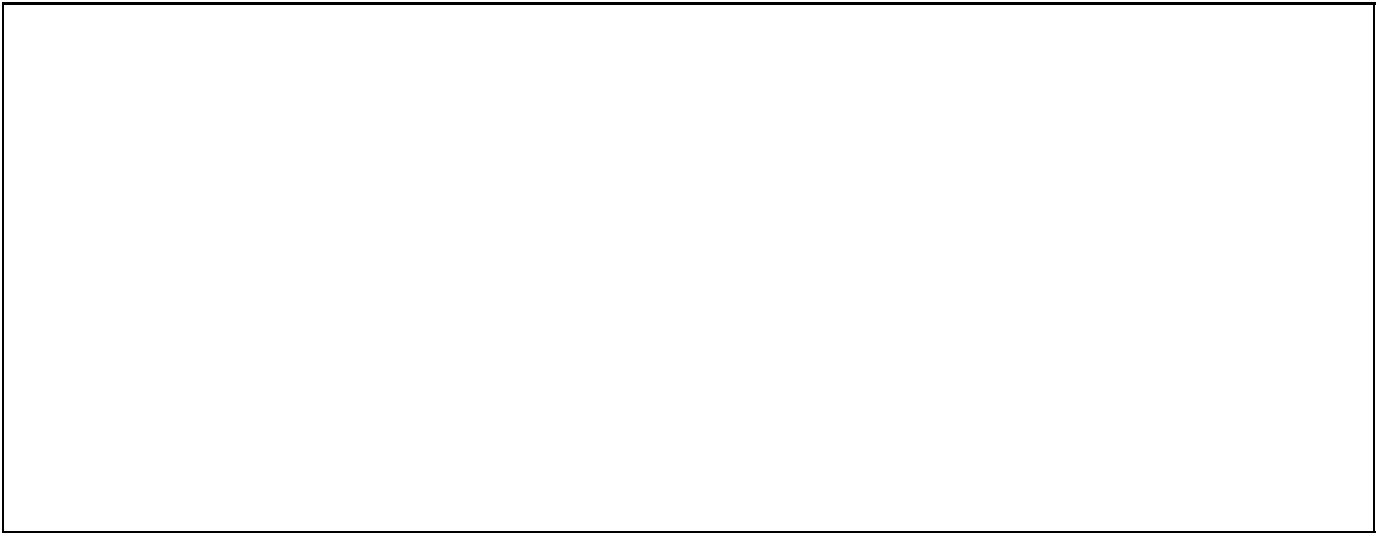

**Fig. 2.** a and b. Isocontours of the difference between 15th and 6th order $X$ and $Y$ transforms between two ESO plates (i.e. ESO 7992 & ESO 8003). Step between isocontour is 0.2 micron in both cases. The first dashed line indicates zero level

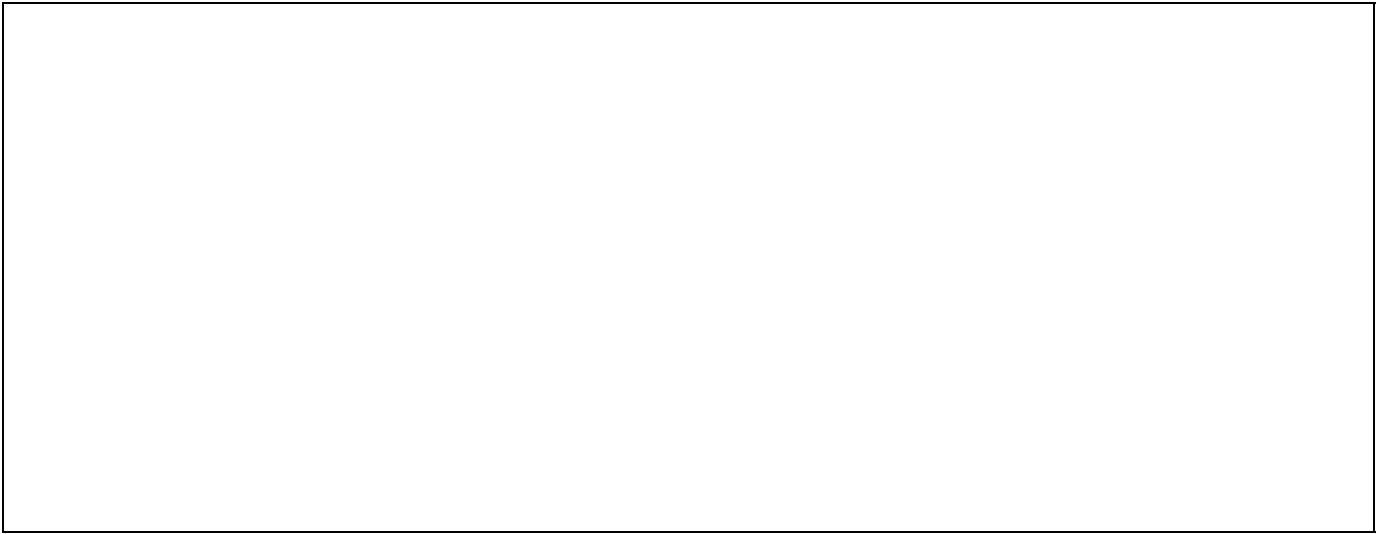

**Fig. 3.** a and b. Isocontours of the difference between 15th and 6th order $X$ and $Y$ transforms between ESO and Palomar plates (i.e. ESO 7992 & POSS 1402,1429). Step between isocontour is 0.4 micron in both cases. The first dashed line indicates zero level

$$x_{B\ cor.} = x_B - [b1 + b2(B-V) + a1 + a2(B-V)] \quad \text{(B filter)} \quad (2)$$

The meaning of equations 1 and 2 is that for ESO V plate, no correction of color was applied but for ESO B plate, the color correction was applied with respect to V plate. Similarly for the POSS copies, we have applied the color correction in the blue filter with respect to red filter, assuming that the color term in the red POSS copy is almost zero in the following way :

$$x_{red\ cor.} = x_{red} \quad \text{(red POSS copy)} \quad (3)$$

For the blue POSS copy :

$$x_{blue\ cor.} = x_{blue} - [b3 + b4(B-V) + a3 + a4(B-V)] \quad (4)$$

Where a3, a4, b3 and b4 are the constants determined by using a linear fit in the residuals of $x$ coordinate on two POSS plates in red and blue filters respectively *versus* B-V color. The total color term in $x$ position is found to be 0.65 $\mu m$ (0.04 arcsecs) per unit in B-V (slope) for the POSS blue plate. The color effect in $y$ coordinate on all the plates was found negligible. After applying the above mentioned color correction, the proper motions of each star have been redetermined.

Figs. 4ab show the residuals in $x$ coordinate on two plates (ESO 7958 and POSS 1402 blue) *versus* B-V indice after color correction.

*4.3. Absolute proper motions*

The mean proper motions of a sample of $\sim 1955$ galaxies uniformly distributed in colors and magnitude ranges have been used to calculate the zero point of the proper motions. The conversion equations obtained are as follows:

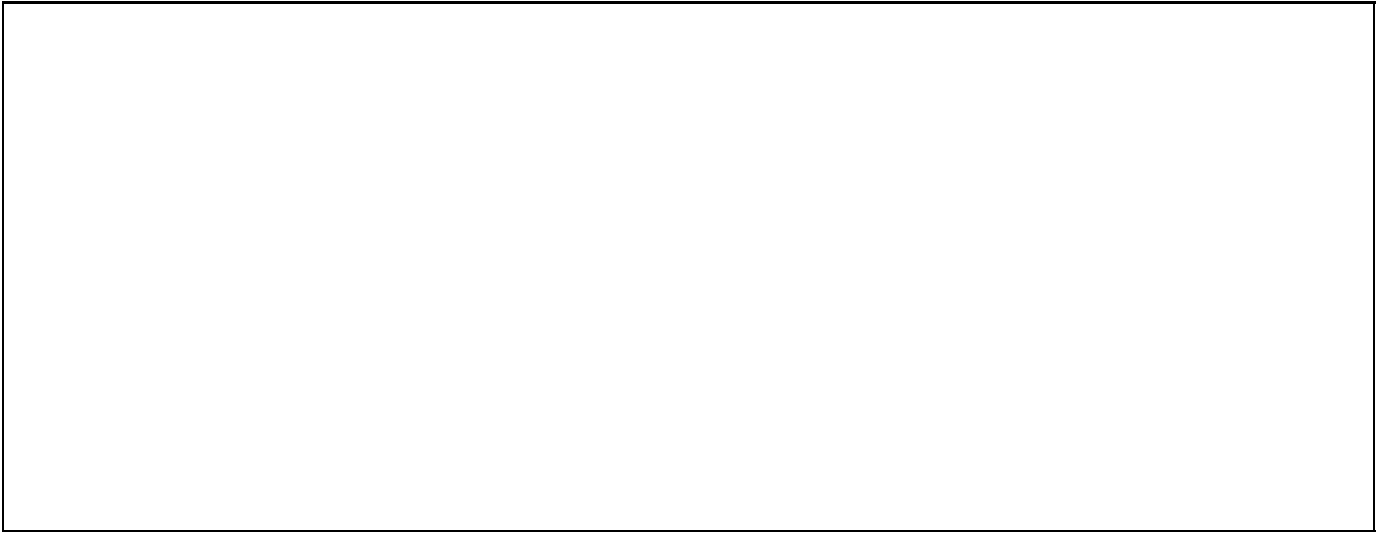

**Fig. 4.** a and b. Residuals in microns in $x$ coordinate as a function of B-V indice for ESO 7958 and POSS 1402 blue plates after color correction

$$\mu_\alpha(abs) = \mu_\alpha - 0.34 \pm 0.03 \quad (''/\text{cen})$$

$$\mu_\delta(abs) = \mu_\delta - 0.28 \pm 0.03 \quad (''/\text{cen})$$

The overall mean random error ($<\sigma_\mu> = \sqrt{\sigma_{\mu x}^2 + \sigma_{\mu y}^2}$) in differential proper motions remains $\sim 0.''3$ /cen to V=17 and is shown in Table 3 as a function of V magnitude.

**Table 3.** The mean error $<\sigma_\mu> = \sqrt{\sigma_{\mu x}^2 + \sigma_{\mu y}^2}$ (arcsec per century) in proper motion as a function of V magnitude

| V mag interval | Number of stars | $<\sigma_\mu>$ (''/cen) |
|---|---|---|
| 9-10 | 49 | 0.60 |
| 10-11 | 147 | 0.50 |
| 11-12 | 384 | 0.39 |
| 12-13 | 823 | 0.33 |
| 13-14 | 1724 | 0.30 |
| 14-15 | 3174 | 0.29 |
| 15-16 | 5510 | 0.30 |
| 16-17 | 8626 | 0.32 |
| 17-18 | 12895 | 0.39 |
| 18-19 | 14615 | 0.52 |

## 5. Discussion

### 5.1. Comparisons with other data

#### 5.1.1. Starcounts

The distribution of the observed starcounts in the V band is shown in Fig 5a. The error bars are $\pm\sqrt{N}$, where N is the number of stars in each bin. To check the quality of the photometric data, we have compared our data set with data from Bienaymé et al. (1992) in the same direction ($l = 3°$, $b = 47°$) and from Gilmore et al. (1985, hereafter referred to as GRH) ($l = 37°$, $b = -51°$). There is a good agreement between the observed total counts from our survey with Bienaymé et al. (1992) counts for $13 \leq V \leq 17$. For $10 \leq V \leq 15$, the agreement is good between GRH and our starcounts. Beyond V=15, GRH counts lie below our counts, which may be explained by the difference in longitude and latitude between the two fields. Comparison of the observed data (present paper) with Besançon model predictions (dashed-dotted line) is also shown in Fig 5a. The observed and model predicted total V counts agree well for $11<V<18$ magnitude bin.

We have also compared starcounts from Besançon model predictions (characteristics described in section 6.1) in the direction of $l = 37°$, $b = -51°$ with GRH data (Fig 5b) and found a good agreement between the observed and model predicted V counts for $12<V<16$ magnitude bin. Beyond V=16, GRH observed counts lie below the model predicted counts. This discrepency may be possible because of the calibration problem in the fainter magnitude part due to small number of standards ($\sim 48$) used in the photographic calibration by GRH.

The mainsprings of Besançon model of population synthesis are described in Robin & Crézé (1986), Bienaymé et al. (1987) and Robin & Oblak (1987).

#### 5.1.2. Color distribution

Figs 6abcd show the comparison of the three data sets in B-V color indice in the overlapping magnitude ranges. There is an excellent agreement between Bienaymé et al. (1992) and our counts for $13 \leq V \leq 17$ magnitude interval. For $13 \leq V \leq 15$, there is a good agreement between GRH data and present survey. Beyond V=15, the data from GRH lie below our distribution, as was already mentioned with the V starcounts.

The comparison of U-B distribution between GRH data and ours is shown in Figs 7abcd. For $11 \leq V \leq 13$, the two distributions are in a good agreement. For $13 \leq V \leq 15$, a shift can be seen between the two data sets. This could be related to unidentified systematic errors with the U photographic calibration. However, there is no shift between the two data sets for faint magnitude intervals.

The observed starcount data are also presented in a tabular form in Table 4 for N(V,B-V) and in Table 5 for N(V,U-B).

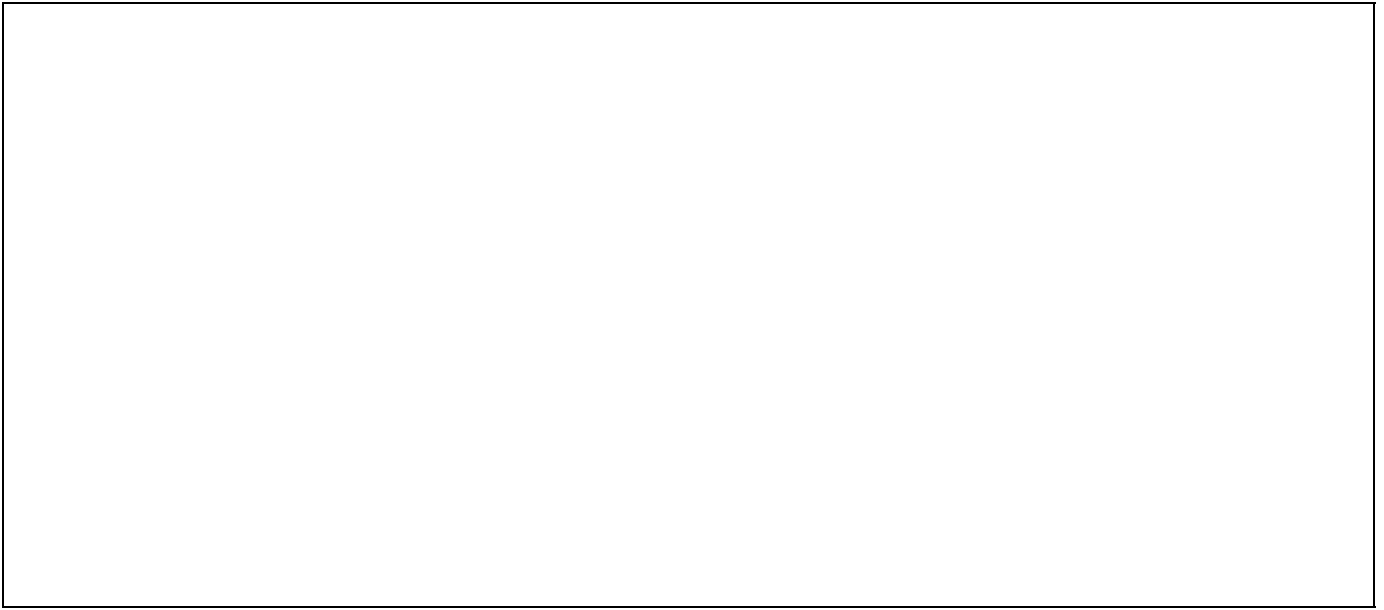

**Fig. 5.** a and b. Comparison of V starcounts with other data sets and with Besançon model

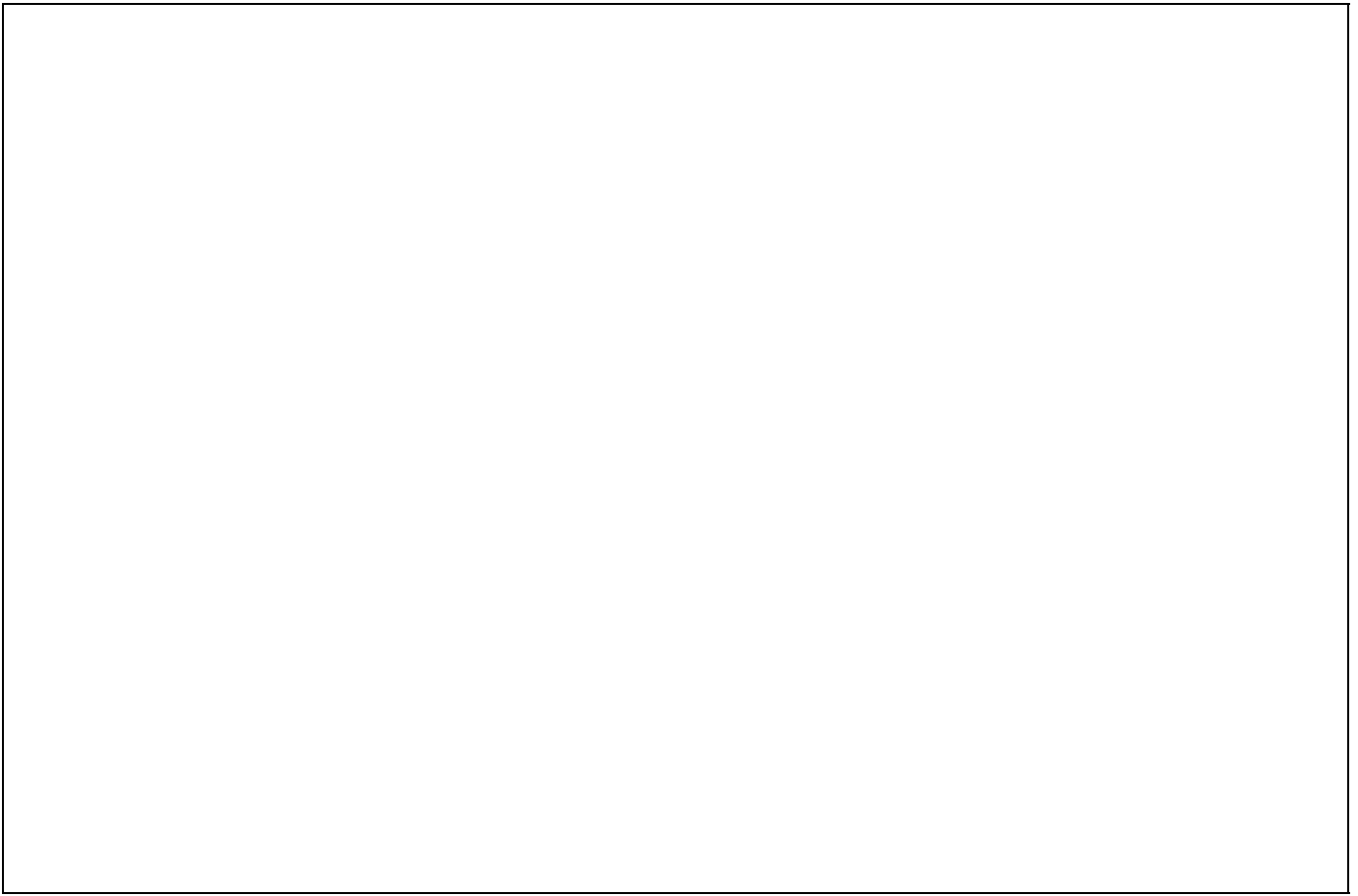

**Fig. 6.** a-d. Comparison of the (B-V) distribution with other data sets. The symbols are as in Fig 5a

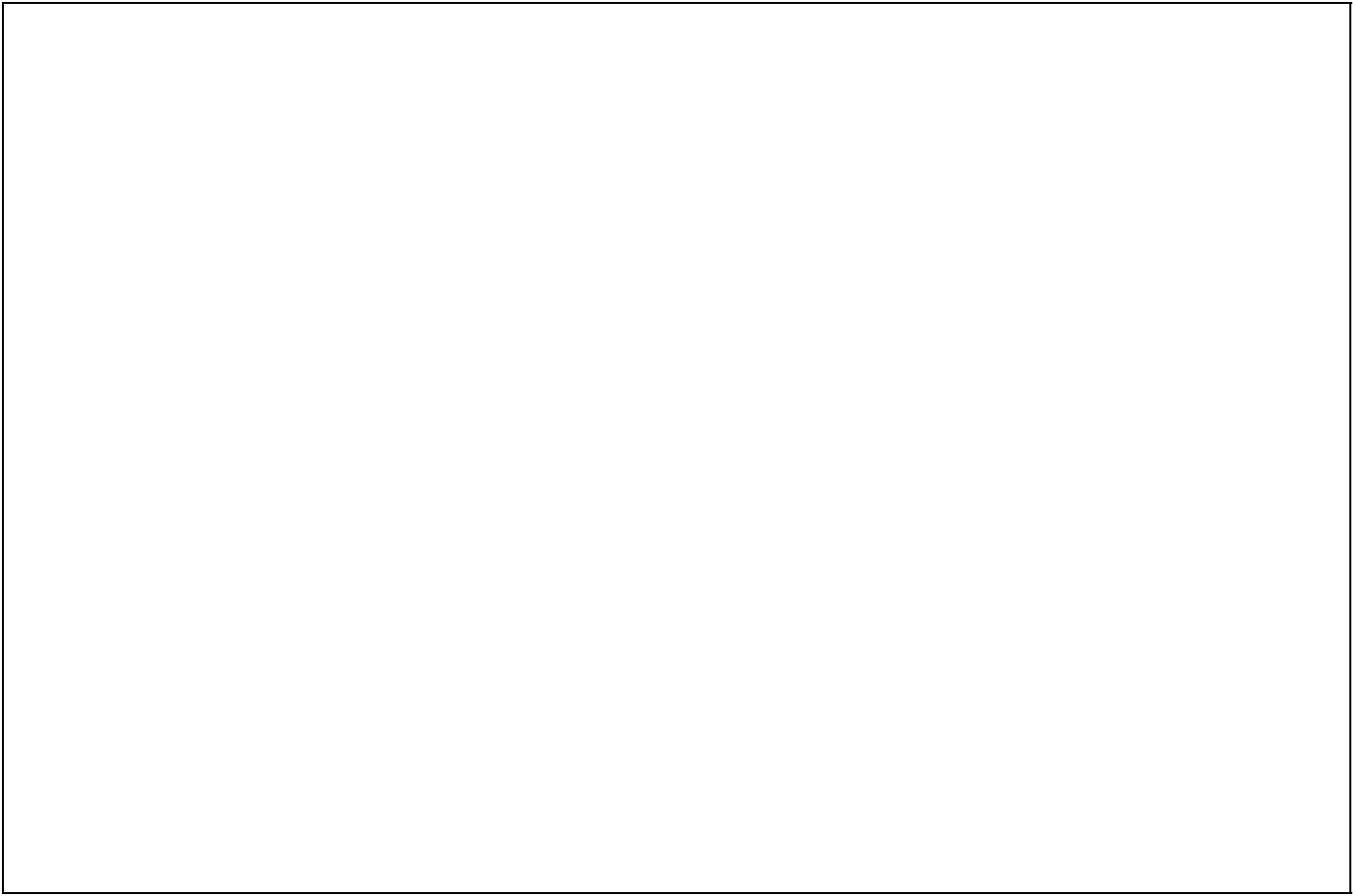

**Fig. 7.** a-d. Comparison of the (U-B) distribution with other data sets. The symbols are as in Fig 5a

**Table 4.** Starcounts over 15.5 square degrees as a function of V and B-V

| B-V<br>V | -0.4 | -0.2 | 0.0 | 0.2 | 0.4 | 0.6 | 0.8 | 1.0 | 1.2 | 1.4 | 1.6 | 1.8 | 2.0 | 2.2 | Total |
|---|---|---|---|---|---|---|---|---|---|---|---|---|---|---|---|
| 10.0-10.5 | 0 | 0 | 0 | 0 | 0 | 3 | 3 | 8 | 6 | 6 | 2 | 0 | 0 | 0 | 28 |
| 10.5-11.0 | 0 | 0 | 0 | 2 | 10 | 23 | 9 | 6 | 7 | 2 | 0 | 0 | 0 | 0 | 59 |
| 11.0-11.5 | 0 | 0 | 0 | 4 | 27 | 28 | 17 | 21 | 9 | 3 | 1 | 0 | 0 | 0 | 110 |
| 11.5-12.0 | 0 | 0 | 0 | 2 | 23 | 54 | 31 | 16 | 6 | 3 | 1 | 0 | 0 | 0 | 136 |
| 12.0-12.5 | 0 | 1 | 0 | 2 | 30 | 85 | 57 | 42 | 12 | 4 | 0 | 0 | 0 | 0 | 232 |
| 12.5-13.0 | 0 | 0 | 1 | 7 | 26 | 108 | 84 | 51 | 12 | 5 | 0 | 0 | 0 | 0 | 294 |
| 13.0-13.5 | 0 | 0 | 1 | 3 | 34 | 173 | 145 | 79 | 16 | 6 | 0 | 0 | 0 | 0 | 457 |
| 13.5-14.0 | 1 | 1 | 3 | 7 | 35 | 247 | 236 | 95 | 21 | 10 | 2 | 0 | 0 | 0 | 658 |
| 14.0-14.5 | 0 | 0 | 1 | 8 | 40 | 287 | 319 | 113 | 31 | 16 | 4 | 0 | 0 | 0 | 819 |
| 14.5-15.0 | 0 | 0 | 1 | 9 | 45 | 374 | 416 | 150 | 51 | 35 | 8 | 1 | 0 | 0 | 1090 |
| 15.0-15.5 | 0 | 0 | 1 | 8 | 62 | 441 | 536 | 237 | 79 | 40 | 14 | 1 | 0 | 1 | 1420 |
| 15.5-16.0 | 0 | 1 | 6 | 9 | 92 | 635 | 753 | 299 | 102 | 74 | 36 | 1 | 1 | 0 | 2009 |
| 16.0-16.5 | 0 | 1 | 5 | 10 | 105 | 766 | 800 | 337 | 175 | 128 | 76 | 10 | 1 | 0 | 2414 |
| 16.5-17.0 | 0 | 4 | 7 | 22 | 180 | 866 | 906 | 453 | 250 | 172 | 109 | 12 | 1 | 0 | 2982 |
| 17.0-17.5 | 2 | 2 | 4 | 25 | 261 | 1059 | 1056 | 544 | 310 | 273 | 138 | 14 | 0 | 0 | 3688 |
| 17.5-18.0 | 0 | 2 | 17 | 51 | 411 | 1167 | 1116 | 637 | 252 | 64 | 0 | 0 | 0 | 0 | 3717 |

### 5.1.3. Proper motions distribution

Figs 8abcd & 9abcd show the distributions of proper motions in $\mu_l$ and $\mu_b$ in different magnitude intervals for Bienaymé et al. (1992) and the present survey. The proper motions distribution in $\mu_l$ and $\mu_b$ agrees well for stars in all the magnitude ranges.

### 5.1.4. Comparison with Guide Star Photometric Catalogue (GSPC) stars

The comparison between 6 photoelectric bright stars measured by Lasker et al. (1988) and our results is shown in Table 6. The agreement is excellent within the accuracy of our photometric measurements.

### 5.1.5. Comparison with PPM stars

Table 7 indicates the differences in the two components of the relative proper motions of PPM stars (Röser & Bastian, 1991) and our results. The agreement is good for bright stars considering the mean error bars in $\mu_\alpha$ and $\mu_\delta$ on the PPM measurements (0.2 to 0.5 arcsec. cen$^{-1}$).

### 5.2. The color-magnitude diagrams (CMDs) of M5 (NGC 5904, C1516+022) and Pal 5 (C1513+000) globular clusters

Figs 10ab show the color-magnitude diagrams (CMDs) of a few cluster member stars for M5 ($\alpha_{1950} = 15^h16.^m1$, $\delta_{1950} = +02°16'$) and Pal 5 ($\alpha_{1950} = 15^h13.^m5$, $\delta_{1950} = +00°05'$) globular clusters measured in our study. The cluster member stars have been identified by cross-matching our catalogue with the catalogues published by Rees (1993) and Smith et al. (1986) for M5 and Pal 5 globular clusters respectively. For the M5 globular cluster (Fig 10a), taking the mean level of the horizontal branch (HB) as V = 15.1±0.03 from the CMD, E(B-V)=0.03±0.02 (Webbink 1985), R = $A_V$/E(B-V)=3.2 and $M_V$(HB)= +0.8±0.15 (Rees 1993), we find a distance modulus of (m-M)$_0$=14.2±0.15, corresponding to a distance d = 6.9±0.48 kpc. Similarly from the CMD of Pal 5, the HB occurs at V=17.43±0.04. With E(B-V)=0.03±0.02 and $M_V$(HB)=+0.6 (Sandage & Hartwick 1977), the distance modulus comes out to be (m-M)$_0$=16.53 or d = 20.27±0.8 kpc. These measured distances for M5 and Pal 5 globular clusters are in good agreement with the determination by Rees (1993) (6.9±0.5 kpc for M5) and Sandage & Hartwick (1977) (21.5 kpc for Pal 5).

## 6. Galactic structure and Kinematics

As described in our previous paper (Ojha et al. 1994), we have selected a subsample of the stars in 0.3≤B-V≤0.9 color interval (mainly F and G- type stars). This type of selection allows to study the different populations of the Galaxy with no kinematical and metallicity biases. However, this is a conservative criterion, but still gives a large sample. The photometric distance of each star has been determined using a $M_V$ and B-V relation. The correction concerning the vertical metallicity gradient ($\frac{\partial [Fe/H]}{\partial z}$ = -0.3 kpc$^{-1}$; Kuijken and Gilmore 1989) is applied on distance measurements. For the distances, a 20% error was estimated from the photometric uncertainties. However, this uncertainty does not include the effect of metallicity.

The U+W and V velocities have been calculated directly from the measured proper motions in $\mu_l$ and $\mu_b$. The algorithm SEM (Stochastic Estimation Maximization; Celeux & Diebolt 1986) is used for the deconvolution of multivariate gaussian distributions. The aim of the SEM algorithm is to resolve the finite mixture density estimation problem under the maximum likelihood approach, using a probabilistic teacher step. Full details can be found in the above mentioned paper. Through SEM one can obtain the number of components of the gaussian mixture (without any assumption on this number), its mean values, dispersions and the percentage of each component with respect to whole sample.

### 6.1. The accuracy of the SEM algorithm

To check the accuracy of the SEM method, we have applied the method on model catalogues (Besançon model predictions) computed towards galactic anticenter direction ($l = 167°$, $b = 47°$). The characteristics of the Besançon model (kinematic part) used in this paper are shown in Table 8. The rotational velocity of the considered population is given by : $V_{rot}$ = $V_{LSR}$+<V>+$V_\odot$ with $V_{LSR}$ = 229 km s$^{-1}$ and $V_\odot$ = 6.3 km s$^{-1}$ in the Besançon model.

In model catalogues, we have a priori knowledge of the intrinsic stellar parameters such as distance and velocities (U, V & W) for each star of different populations. The same selection criterion in B-V as described above (see section 6) has been applied in model catalogues. Fig 11a shows the distribution of stars (1000≤r≤1500 pc) in V velocity (solid line) and the distribution of each population (dashed line- disk, dotted line-thick disk, dashed-dotted line-halo) predicted by the model in the B-V interval of 0.3 to 0.9. The halo population is only 1% of the whole population in this distance bin.

Table 9 shows the comparison of percentage of each population obtained from SEM algorithm (applied to a model catalogue) and by model predictions itself for different distance (line of sight) intervals. The percentage of each population obtained from SEM method and by model predictions is in a good agreement except for the bin 1500≤r≤2000 pc where SEM solution is less stable. This may be due to varying contribution of different populations, where one population dominates strongly the other populations. In that case SEM is not able to deconvolve the 2 populations clearly, because one of the population contains few high velocity stars. Fig 11b shows the distribution of stars (1500≤r≤2000 pc) in V velocity (solid line) and the distribution of each population (dashed line- disk, dotted line-thick disk, dashed-dotted line-halo) predicted by the model in the B-V interval of 0.3 to 0.9. The last column in Table 9 shows the measurements of the asymmetric drift for the thick disk population derived by SEM algorithm from model catalogue, which depends on the value of circular velocity for the thick disk adopted for model predictions.

In Figs. 12ab, the three gaussian populations representing the thin disk, the thick disk and the halo are overplotted on the V velocity histogram for the distance intervals 1000≤r≤1500 pc and 1500≤r≤2000 pc. By comparing Figs 11a & 12a, we find that the deconvolution of the multivariate gaussian distributions using SEM algorithm is in a good agreement with the model predictions. However, in the distance interval 1500≤r≤2000 pc, the SEM algorithm is not able to separate clearly the gaussian components. The difference can be seen by comparing Figs 11b and 12b. However, for the farthest distances (r≥2000 pc), the SEM algorithm works nicely (see Table 9).

Table 5. Starcounts over 15.17 square degrees as a function of V and U-B

| U-B<br>V | -1.0 | -0.8 | -0.6 | -0.4 | -0.2 | 0.0 | 0.2 | 0.4 | 0.6 | 0.8 | 1.0 | 1.2 | 1.4 | 1.6 | 1.8 | 2.0 | 2.2 | Total |
|---|---|---|---|---|---|---|---|---|---|---|---|---|---|---|---|---|---|---|
| 10.0-10.5 | 0 | 0 | 0 | 0 | 1 | 0 | 1 | 2 | 0 | 1 | 2 | 1 | 2 | 3 | 1 | 2 | 2 | 18 |
| 10.5-11.0 | 0 | 0 | 1 | 1 | 6 | 15 | 8 | 2 | 1 | 4 | 2 | 4 | 1 | 1 | 2 | 1 | 0 | 49 |
| 11.0-11.5 | 0 | 0 | 0 | 1 | 12 | 24 | 15 | 9 | 10 | 5 | 10 | 5 | 6 | 2 | 2 | 1 | 3 | 105 |
| 11.5-12.0 | 0 | 0 | 1 | 1 | 4 | 21 | 34 | 18 | 12 | 5 | 5 | 6 | 3 | 1 | 4 | 2 | 0 | 117 |
| 12.0-12.5 | 0 | 0 | 0 | 1 | 10 | 19 | 46 | 32 | 36 | 20 | 21 | 10 | 6 | 6 | 1 | 2 | 0 | 210 |
| 12.5-13.0 | 0 | 0 | 0 | 0 | 9 | 41 | 58 | 56 | 37 | 29 | 18 | 5 | 7 | 5 | 1 | 1 | 0 | 267 |
| 13.0-13.5 | 0 | 0 | 0 | 0 | 13 | 65 | 118 | 79 | 48 | 43 | 24 | 16 | 5 | 4 | 1 | 1 | 0 | 417 |
| 13.5-14.0 | 0 | 1 | 1 | 1 | 18 | 116 | 164 | 121 | 81 | 55 | 30 | 16 | 5 | 0 | 2 | 0 | 1 | 612 |
| 14.0-14.5 | 0 | 1 | 1 | 0 | 26 | 169 | 233 | 139 | 89 | 37 | 33 | 21 | 5 | 0 | 0 | 1 | 0 | 755 |
| 14.5-15.0 | 0 | 0 | 0 | 2 | 53 | 240 | 269 | 172 | 112 | 61 | 38 | 41 | 15 | 2 | 2 | 0 | 0 | 1022 |
| 15.0-15.5 | 0 | 0 | 0 | 6 | 92 | 366 | 294 | 207 | 136 | 83 | 56 | 42 | 15 | 0 | 0 | 0 | 0 | 1297 |
| 15.5-16.0 | 1 | 0 | 0 | 11 | 219 | 529 | 409 | 268 | 160 | 110 | 42 | 6 | 0 | 0 | 0 | 0 | 0 | 1755 |
| 16.0-16.5 | 0 | 1 | 3 | 48 | 344 | 629 | 394 | 250 | 127 | 51 | 7 | 0 | 0 | 0 | 0 | 0 | 0 | 1854 |
| 16.5-17.0 | 5 | 8 | 9 | 99 | 541 | 688 | 383 | 144 | 19 | 0 | 1 | 0 | 0 | 0 | 0 | 0 | 0 | 1897 |
| 17.0-17.5 | 23 | 28 | 21 | 234 | 713 | 469 | 92 | 4 | 1 | 0 | 0 | 0 | 0 | 0 | 0 | 0 | 0 | 1585 |
| 17.5-18.0 | 22 | 26 | 50 | 235 | 183 | 42 | 2 | 1 | 0 | 0 | 0 | 0 | 0 | 0 | 0 | 0 | 0 | 561 |

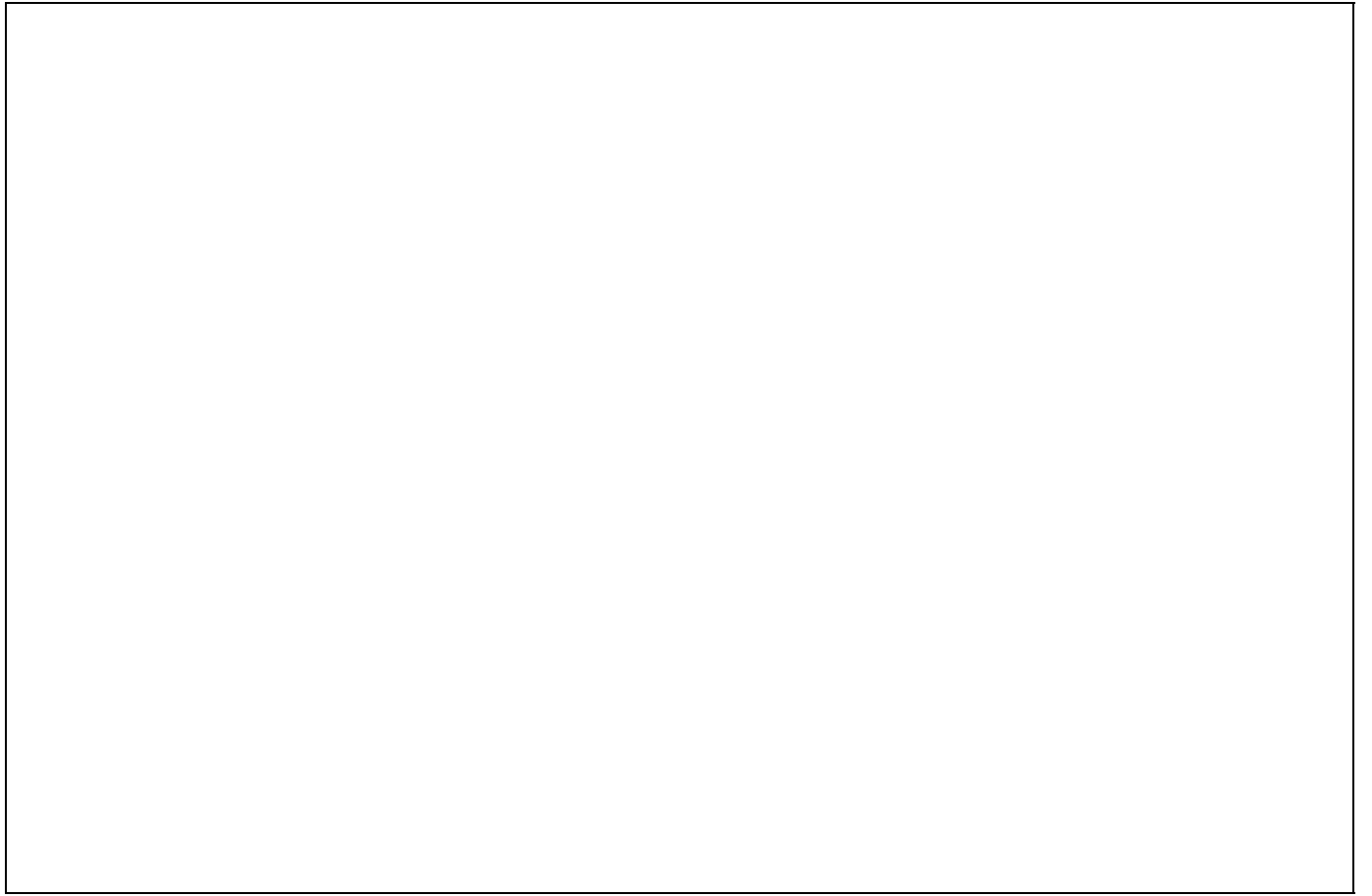

Fig. 8. a-d. $\mu_l$ observed proper motion distributions for four magnitude intervals for Bienaymé et al. (1992) and present survey. Symbols are as in Fig 5a

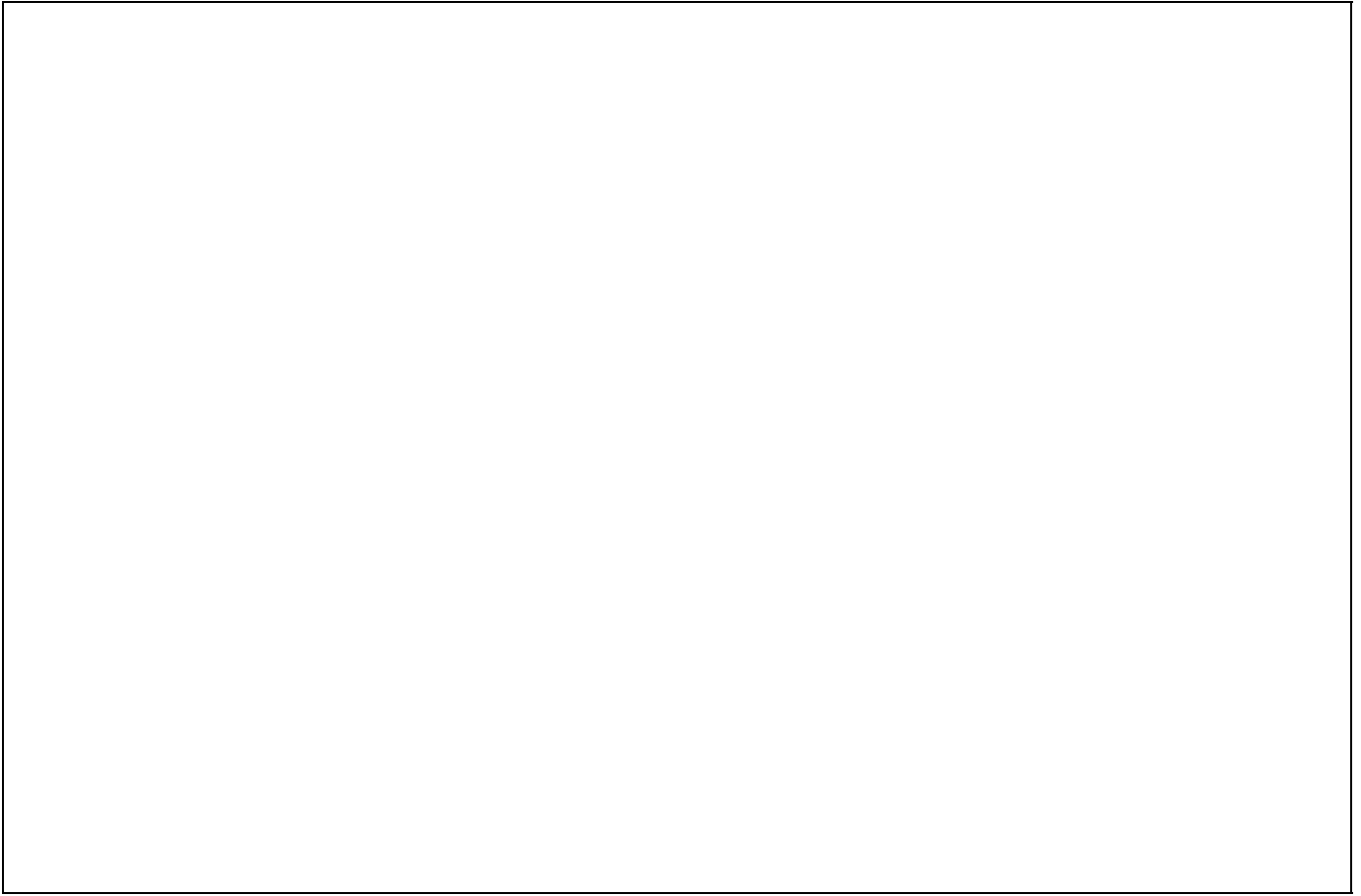

**Fig. 9.** a-d. $\mu_b$ observed proper motion distributions for four magnitude intervals for Bienaymé et al. (1992) and present survey. Symbols are as in Fig 5a

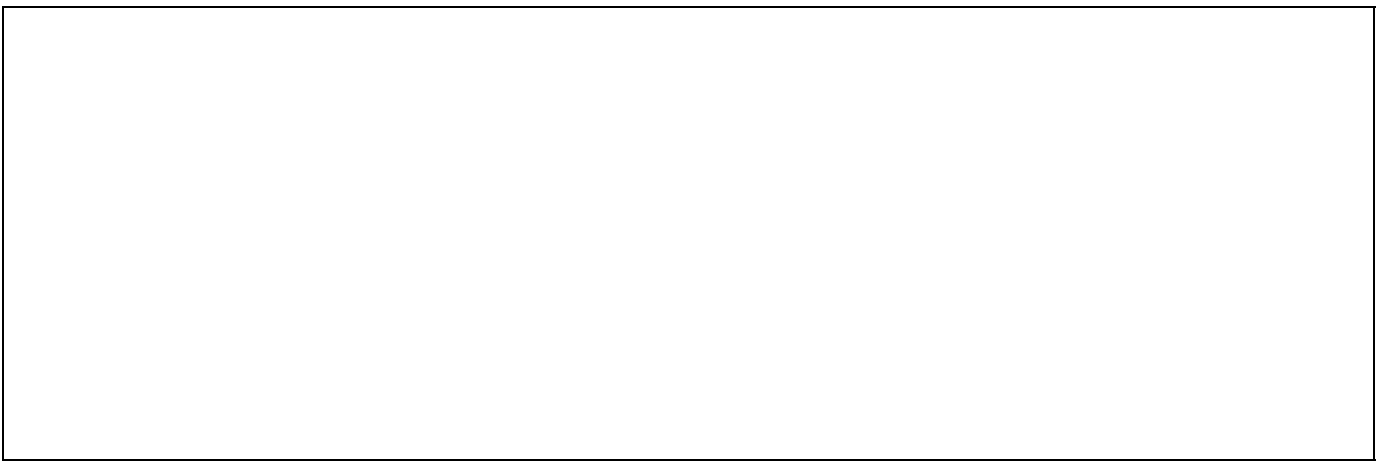

**Fig. 10.** a and b. Color-magnitude diagrams of M5 and Pal 5 globular clusters

**Table 6.** Comparison of photographic magnitudes with GSPC photoelectric magnitudes

| Stars | $V_{GSPC}$ | $V_{M5}$ | $(B-V)_{GSPC}$ | $(B-V)_{M5}$ |
|---|---|---|---|---|
| S869-B | $9.31 \pm 0.02$ | $9.25 \pm 0.07$ | $1.38 \pm 0.00$ | $1.32 \pm 0.07$ |
| S869-C | $11.00 \pm 0.01$ | $11.01 \pm 0.08$ | $0.40 \pm 0.00$ | $0.33 \pm 0.05$ |
| S869-D | $11.67 \pm 0.00$ | $11.73 \pm 0.09$ | $0.81 \pm 0.01$ | $0.78 \pm 0.06$ |
| S869-E | $13.15 \pm 0.00$ | $13.16 \pm 0.09$ | $0.65 \pm 0.00$ | $0.65 \pm 0.07$ |
| S869-F | $13.99 \pm 0.03$ | $14.01 \pm 0.09$ | $0.74 \pm 0.04$ | $0.77 \pm 0.06$ |
| S869-G | $15.04 \pm 0.03$ | $14.98 \pm 0.09$ | $0.62 \pm 0.03$ | $0.73 \pm 0.08$ |

**Table 7.** Comparison with PPM high proper motion stars. M5 corresponds to the present paper

| PPM Number | $\mu_\alpha$ M5 ("/cen) | $\mu_{\alpha err}$ M5 ("/cen) | $\mu_\delta$ M5 ("/cen) | $\mu_{\delta err}$ M5 ("/cen) | $(\mu_{\alpha PPM} - \mu_{\alpha M5})$ ("/cen) | $(\mu_{\delta PPM} - \mu_{\delta M5})$ ("/cen) | V |
|---|---|---|---|---|---|---|---|
| 179575 | -0.27 | 0.42 | -1.09 | 0.77 | 0.07 | -0.61 | 9.0 |
| 161550 | -1.47 | 0.35 | 0.42 | 0.47 | -0.37 | 0.48 | 10.1 |
| 161539 | 1.30 | 0.42 | -1.83 | 0.88 | 0.10 | 0.63 | 8.7 |
| 161521 | 0.84 | 0.16 | -0.19 | 0.28 | 0.86 | -0.91 | 10.9 |
| 179560 | -1.10 | 0.21 | -1.34 | 0.51 | -0.30 | -0.16 | 9.2 |
| 161486 | -1.22 | 0.38 | -0.04 | 0.42 | 0.02 | 0.23 | 9.9 |
| 161475 | -2.49 | 0.58 | 1.97 | 0.33 | 0.39 | 0.13 | 10.5 |
| 161450 | 5.84 | 0.59 | -1.93 | 0.32 | -0.74 | 0.83 | 8.6 |
| 161442 | -2.38 | 0.36 | -0.45 | 0.37 | -0.62 | 0.75 | 10.3 |
| 161438 | 2.98 | 0.49 | -1.39 | 1.05 | -0.39 | -0.81 | 9.3 |
| 161436 | -3.23 | 0.36 | -0.11 | 0.54 | -0.07 | -0.19 | 9.1 |
| 161395 | -0.77 | 0.26 | 1.84 | 0.54 | -0.73 | -0.74 | 9.2 |
| 161377 | 1.15 | 1.81 | 0.58 | 2.01 | -0.85 | 0.62 | * |
| 161364 | -0.22 | 0.25 | -2.16 | 0.28 | -0.38 | -0.14 | 10.1 |
| 161345 | 1.42 | 0.34 | -3.63 | 0.30 | -0.32 | -0.17 | 9.9 |
| 179503 | -1.28 | 0.40 | -1.35 | 1.80 | 0.78 | 0.75 | * |
| 161327 | -2.63 | 1.06 | -3.15 | 0.79 | 0.33 | 0.35 | 8.9 |

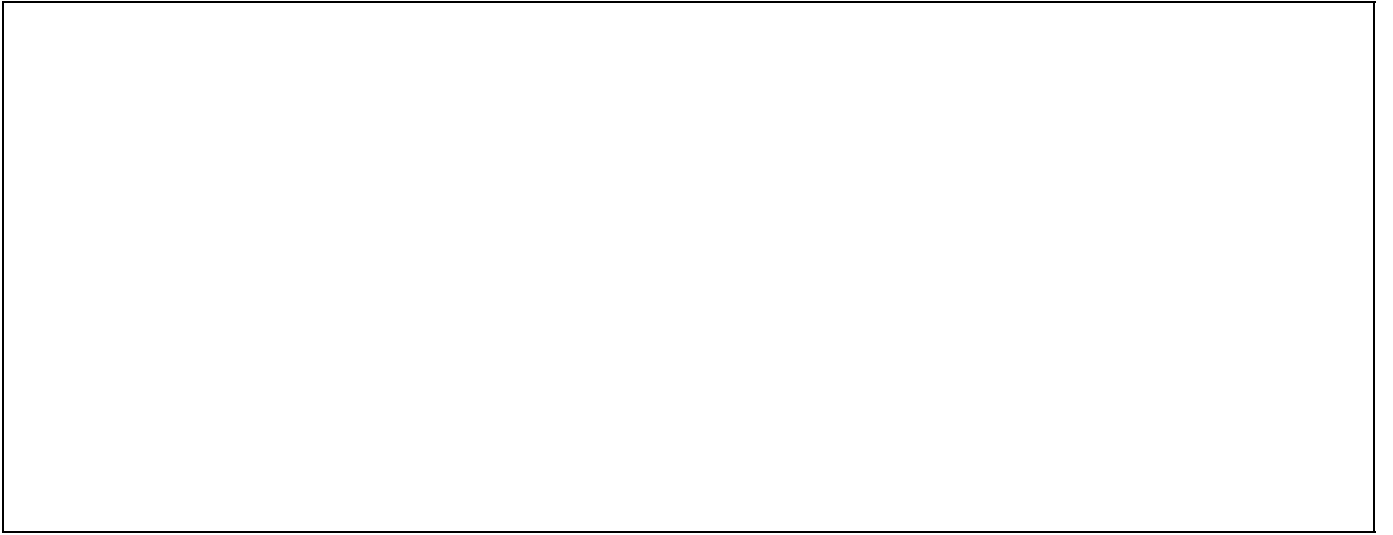

**Fig. 11.** a and b. Predicted V velocity of stars in distance intervals $1000 \leq r \leq 1500$ pc and $1500 \leq r \leq 2000$ pc towards $l = 167°$, $b = 47°$ in the color range 0.3<B-V<0.9. Full line– all stars; dashed line–disk; dotted line–thick disk; dashed-dotted– halo

Table 8. The characteristics of Besançon model used for the comparison with real data

| Disk | Age (in yrs) | $\sigma_U$ km s$^{-1}$ | $\sigma_V$ km s$^{-1}$ | $\sigma_W$ km s$^{-1}$ | Vrot km s$^{-1}$ | [Fe/H] | $\frac{\partial [Fe/H]}{\partial z}$ dex kpc$^{-1}$ | $\frac{\partial [Fe/H]}{\partial r}$ dex kpc$^{-1}$ |
|---|---|---|---|---|---|---|---|---|
| | 0-.15E9 | 16.7 | 10.8 | 6.0 | - | 0.01 | -0.5 | -0.07 |
| | .15-1E9 | 19.8 | 12.8 | 10.0 | - | 0.03 | -0.5 | -0.07 |
| | 1-2E9 | 27.2 | 17.6 | 13.0 | - | 0.03 | -0.5 | -0.07 |
| | 2-3E9 | 30.2 | 19.5 | 18.5 | - | 0.01 | -0.5 | -0.07 |
| | 3-5E9 | 34.5 | 22.2 | 20.0 | - | -0.07 | -0.5 | -0.07 |
| | 5-7E9 | 34.5 | 22.2 | 20.0 | - | -0.14 | -0.5 | -0.07 |
| | 7-10E9 | 34.5 | 22.2 | 20.0 | - | -0.37 | -0.5 | -0.07 |
| Thick disk | - | 51.0 | 38.0 | 35.0 | 190 | -0.7 | 0.0 | 0.0 |
| Halo | - | 131.0 | 106.0 | 85.0 | 0 | -1.7 | 0.0 | 0.0 |

Table 9. Comparison of the results from SEM algorithm applied to a catalogue of predictions obtained with Besançon model towards galactic anticenter direction ($l = 167°$, $b = 47°$) for the three populations of the Galaxy. $<V>_{TD} = V_{rot}-V_{LSR}-V_\odot$ is the mean V velocity for the thick disk stars. The standard error on $<V>$ is $\frac{\sigma_V}{\sqrt{N}}$, where N is the total number of stars in each distance bin

| r (pc) | Besançon model | | | SEM | | | $<V>_{TD}$ (km sec$^{-1}$) |
|---|---|---|---|---|---|---|---|
| | % D | % TD | % H | % D | % TD | % H | |
| 0≤r≤500 | 89.0 | 10.8 | 0.2 | 79.8 | 19.8 | 0.4 | -35.4±1 |
| 500≤r≤1000 | 71.0 | 28.5 | 0.5 | 68.7 | 30.8 | 0.5 | -43.1±1 |
| 1000≤r≤1500 | 45.3 | 53.5 | 1.2 | 47.8 | 51.0 | 1.2 | -46.2±1 |
| 1500≤r≤2000 | 18.8 | 78.7 | 2.5 | 38.2 | 59.5 | 2.3 | -53.2±1 |
| 2000≤r≤2500 | 2.7 | 93.2 | 4.1 | 0.0 | 96.5 | 3.5 | -45.7±2 |
| 2500≤r≤3500 | 0.0 | 90.2 | 9.8 | 0.0 | 88.9 | 11.1 | -48.5±2 |
| 3500≤r≤10000 | 0.0 | 65.6 | 34.4 | 0.0 | 66.8 | 33.2 | -47.0±3 |

*6.2. The kinematical parameters of the thin disk*

Applying the SEM algorithm on the real sample, a population with a low velocity dispersion corresponding to the thin disk, has been identified in 4 bins of distance up to 1.2 kpc above the plane. Table 10 shows the new estimates of the kinematical parameters of the thin disk population derived from SEM algorithm. A modest vertical gradient can be seen in the asymmetric drift and in velocity dispersions ($\sigma_V$ & $\sigma_{U+W}$), and can be interpreted as the superposition of young and old disk populations. The weighted averages of the estimated parameters are: $<U+W> = 3\pm1$ km sec$^{-1}$, $<V> = -13\pm1$ km sec$^{-1}$, $\sigma_{U+W} = 35\pm1$ km sec$^{-1}$ and $\sigma_V = 28\pm1$ km sec$^{-1}$. The $<U+W>$ and $<V>$ velocities are relative to the Sun.

*6.3. The kinematical parameters of the thick disk*

Table 11 shows the kinematical parameters derived for the thick disk population. In the distance bins (z) 1591 and 2121 pc, the SEM algorithm is not able to deconvolve the 2 populations clearly. This effect was explained in section 6.1. The weighted averages of the estimated parameters are: $<U+W> = 4\pm3$ km sec$^{-1}$, $<V> = -49\pm3$ km sec$^{-1}$, $\sigma_{U+W} = 66\pm3$ km sec$^{-1}$ and $\sigma_V = 56\pm2$ km sec$^{-1}$.

6.3.1. The asymmetric drift of the thick disk

From our study, the mean value of the asymmetric drift for the thick disk population comes out to be 38±3 km sec$^{-1}$ with respect to LSR (Local Standard of Rest), which is similar to other recent results : 41±16 km sec$^{-1}$ (Soubiran 1993), 46±4 km sec$^{-1}$ (Ojha et al. 1994). The rotational velocity of the LSR is $V_{LSR} = 220$ km s$^{-1}$ (IAU 1985) and $V_\odot = 11$ km s$^{-1}$ (Delhaye 1965).

The measured values of the asymmetric drift have ranged from 20 km s$^{-1}$ (Norris 1987) to 100 km s$^{-1}$ (Wyse & Gilmore 1986). Recently, Majewski (1992) finds that a mean velocity of the Intermediate Populations II varying from -24±6 km s$^{-1}$ near 700 pc to -122±26 km s$^{-1}$ at z∼6-7 kpc. Fig 13 presents the asymmetric drift measurements of the thick disk population from selected studies as a function of z distances. From the figure, no clear dependence with z is found in the asymmetric drift measurements for the thick disk population. However, the gradient could be seen in velocity if *no separation* was made between the 3 populations and is shown in Fig 13 as dotted-dashed line (h) (Soubiran 1993, north pole); dashed line (g) (Ojha et al. 1994, anticenter field) and dotted line (f) (present papar) respectively.

*6.4. The kinematical parameters of the halo*

Since our survey is not so deep enough to contain much of the halo population. The error bars are larger in the farthest distance bins (z≥2.1 kpc) and the estimates of the halo parameters are less certain in these distance bins. However, the kinematics of the halo population has been deduced from the SEM deconvolution in the 3 highest distance bins (between z = 1.2 to 2.0 kpc). The weighted averages of the estimated parameters are : $<U+W> = 6\pm10$ km s$^{-1}$, $<V> = -144\pm9$ km s$^{-1}$, $\sigma_{U+W} = 156\pm7$ km s$^{-1}$ and $\sigma_V = 145\pm7$ km s$^{-1}$. The dispersion in the V velocity ($\sigma_V$) is higher in our case then other determinations : 96±8 km s$^{-1}$ (Carney & Latham 1986), 106±6 km s$^{-1}$ (Norris 1986), 96±13 km s$^{-1}$ (Morrison et al.

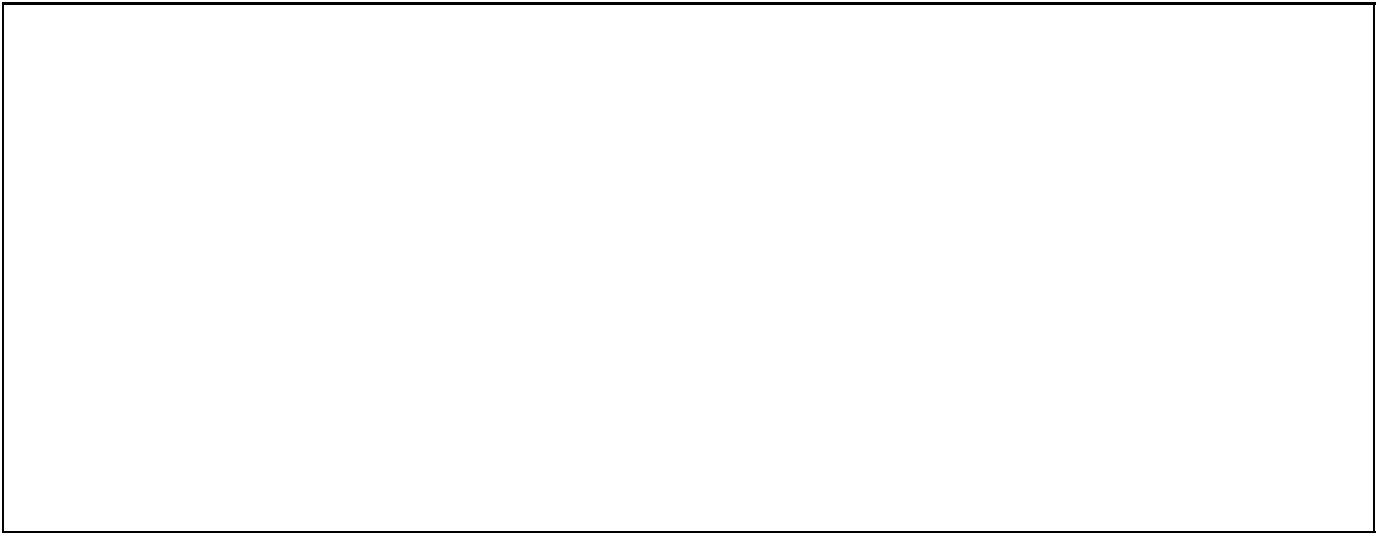

**Fig. 12.** a and b. Histograms of the V velocity for stars in distance intervals 1000≤r≤1500 pc and 1500≤r≤2000 pc towards galactic anticenter direction ($l = 167°$, $b = 47°$). The 3 gaussian components solution of SEM corresponding to the thin disk, the thick disk and the halo are overplotted. The halo is only 1% of the whole population in 1000≤r≤1500 pc distance bin

1990), 100 km s$^{-1}$ (Ryan & Norris 1991) and 79±6 km s$^{-1}$ (Soubiran 1993). This high dispersion in V velocity is observed in the 3 distance bins and could indicate a contamination by the thick disk stars. This approach does not allow to account for the halo specific metallicity, which is also a source of over-estimation of the distance and velocity for the halo stars. We have estimated that for halo stars, there may be an error in our distances or velocities as large as 30 % (up to $z = 2.1$ kpc) because of the metallicity-dependent and photometric errors.

With a mean value of the asymmetric drift of 133±9 km s$^{-1}$ with respect to the LSR, the halo is found to have a *prograde* rotational velocity of 87±9 km s$^{-1}$. Most of the previous determinations give a rotational velocity of the halo ranging between 0 and 60 km s$^{-1}$ : 37±10 km s$^{-1}$ (Norris 1986), 25±15 km s$^{-1}$ (Morrison et al. 1990), 30±10 km s$^{-1}$ (Ryan & Norris 1991), 58±12 km s$^{-1}$ (Soubiran 1993), while Reid (1990) and Majewski (1992) obtained an intrinsically different results from their deep proper motion surveys with a *retrograde* rotation of 30 and 55 km s$^{-1}$ respectively.

**Table 10.** The kinematical parameters of the thin disk (km sec$^{-1}$) derived from SEM algorithm. The standard errors on <U+W>, <V>, $\sigma_{U+W}$ and $\sigma_V$ are $\frac{\sigma_{U+W}}{\sqrt{N}}$, $\frac{\sigma_V}{\sqrt{N}}$, $\frac{\sigma_{U+W}}{\sqrt{2N}}$ and $\frac{\sigma_V}{\sqrt{2N}}$ respectively. Where N is the number of stars in each distance bin and the velocities are relative to the Sun

| $<z>$ (pc) | 177 | 530 | 884 | 1237 |
|---|---|---|---|---|
| $N_{tot}$ | 840 | 2121 | 2172 | 2208 |
| $p(\%)_{Disk}$ | 85±17 | 80±2 | 62±12 | 56±10 |
| <U+W> | -4±1 | 5±1 | 3±1 | 8±2 |
| <V> | -11±1 | -12±1 | -12±1 | -17±2 |
| $\sigma_{U+W}$ | 23±1 | 33±1 | 38±1 | 45±2 |
| $\sigma_V$ | 18±1 | 25±1 | 32±1 | 35±2 |

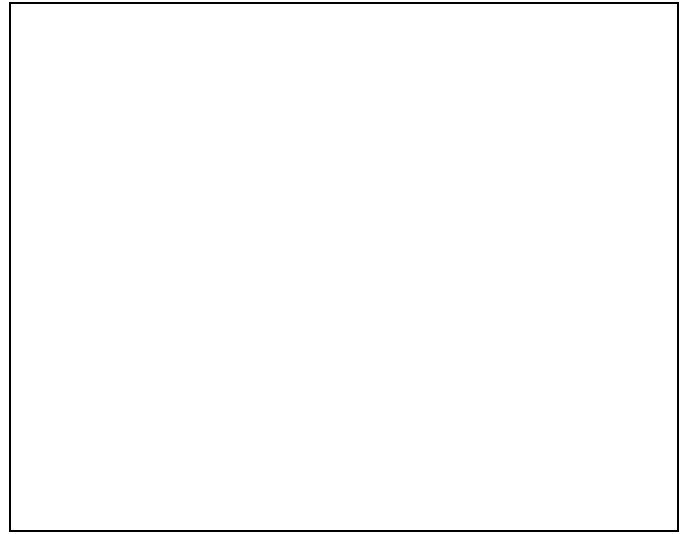

**Fig. 13.** The measured asymmetric drift of the thick disk population, plotted as a function of z distances from the proper motion selected samples (open circle): a– present survey; (open square): b– anticenter field (Ojha et al. 1994); (open diamond): c– North Pole (Soubiran 1993). Solid line (e) represents Majewski's (1992) survey for only *intermediate population* stars. The dotted (f), dashed (g) and dashed-dotted (h) lines represent our present survey, anticenter field (Ojha et al. 1994) and north pole (Soubiran 1993) with *no separation* between the three populations. Vertical bars indicate the error $\frac{\sigma_V}{\sqrt{N}}$ in V velocity. Where N is the number of stars in each distance bin

## 7. Conclusion

We have tested upon the salient features of our data set at the intermediate galactic latitude and their implications on the kinematic properties of the Galaxy. The old disk is found with mean velocity (U+W,V) = (3±1,-13±1) km sec$^{-1}$ and velocity dispersions ($\sigma_{U+W}$,$\sigma_V$) = (35±1,28±1) km sec$^{-1}$. The thick disk is found with asymmetric drift of 49±3 km sec$^{-1}$ and velocity dispersions $(\sigma_{U+W}, \sigma_V) = (66\pm3, 56\pm2)$ km sec$^{-1}$ with respect to the Sun. No dependence with z is found in the asymmetric drift of the thick disk population up to distance of z = 3.4 kpc. The halo is found to have a *prograde* rotation. Additional informations will be gained when other surveys in different directions of the Galaxy will be analyzed globally by using Besançon model.

Table 11. Kinematical parameters for the thick disk population (same as in Table 10)

| $<z>$(pc) | 177 | 530 | 884 | 1237 | 1591 | 2121 | 3359 |
|---|---|---|---|---|---|---|---|
| $N_{tot}$ | 840 | 2121 | 2172 | 2208 | 2060 | 2343 | 1092 |
| $p(\%)_{TD}$ | 15±10 | 20±1 | 31±11 | 36±9 | 85±6 | 64±4 | 55±9 |
| $<U+W>$ | 10±4 | 7±4 | 18±3 | 6±3 | 7±2 | -6±2 | -12±5 |
| $<V>$ | -32±4 | -49±4 | -46±2 | -53±2 | -32±2 | -31±2 | -56±4 |
| $\sigma_{U+W}$ | 38±3 | 77±3 | 64±2 | 77±2 | 65±1 | 75±2 | 109±3 |
| $\sigma_V$ | 43±3 | 69±3 | 49±2 | 61±2 | 55±1 | 61±1 | 93±3 |

*Acknowledgements.* This work forms a part of the PhD thesis of DKO, which was partially supported by the Indo-French Center for the Promotion of Advanced Research / Centre Franco-Indien Pour la Promotion de la Recherche Avancée, New-Delhi (India). We thank Dr. Neill Reid for letting us use their UBV catalogue. We especially thank referee Dr. Gerry Gilmore for his comments. We also thank all the MAMA, OCA, ESO and Leiden Observatory staffs who made this investigation possible.